\documentclass[draft,journal,a4paper,oneside,onecolumn]{IEEEtran} 
\usepackage{amsmath,amsfonts, amssymb}

\newtheorem{thm}{Theorem}
\newtheorem{cor}[thm]{Corollary}
\newtheorem{lem}{Lemma}
\newtheorem{df}{Definition}
\newtheorem{rem}{Remark}

\newcommand{\markov}{\leftrightarrow}
\newcommand{\fMAP}{f_{\mathrm{MAP}}}
\newcommand{\GFq}{\mathrm{GF}(q)}
\newcommand{\A}{\mathcal{A}}
\newcommand{\B}{\mathcal{B}}
\newcommand{\C}{\mathcal{C}}
\newcommand{\G}{\mathcal{G}}
\newcommand{\cH}{\mathcal{H}}
\newcommand{\M}{\mathcal{M}}
\newcommand{\cP}{\mathcal{P}}
\newcommand{\cS}{\mathcal{S}}
\newcommand{\T}{\mathcal{T}}
\newcommand{\U}{\mathcal{U}}
\newcommand{\V}{\mathcal{V}}
\newcommand{\X}{\mathcal{X}}
\newcommand{\Y}{\mathcal{Y}}
\newcommand{\hcH}{\widehat{\mathcal H}}
\newcommand{\hu}{\widehat{u}}
\newcommand{\hU}{\widehat{U}}
\newcommand{\tX}{\widetilde{X}}
\newcommand{\aalpha}{\boldsymbol{\alpha}}
\newcommand{\bbeta}{\boldsymbol{\beta}}
\newcommand{\cc}{\boldsymbol{c}}
\newcommand{\mm}{\boldsymbol{m}}
\newcommand{\bp}{\boldsymbol{p}}
\newcommand{\bt}{\boldsymbol{t}}
\newcommand{\uu}{\boldsymbol{u}}
\newcommand{\xx}{\boldsymbol{x}}
\newcommand{\yy}{\boldsymbol{y}}
\newcommand{\WW}{\boldsymbol{W}}
\newcommand{\XX}{\boldsymbol{X}}
\newcommand{\YY}{\boldsymbol{Y}}
\newcommand{\vphi}{\varphi}
\newcommand{\e}{\varepsilon}
\newcommand{\sfA}{\mathsf{A}}
\newcommand{\sfB}{\mathsf{B}}
\newcommand{\osfA}{\overline{\sfA}}
\newcommand{\sfcc}{\boldsymbol{\mathsf{c}}}
\newcommand{\lrB}[1]{\left[{#1}\right]}
\newcommand{\lrb}[1]{\left\{{#1}\right\}}
\newcommand{\lrsb}[1]{\left({#1}\right)}
\newcommand{\lrbar}[1]{\left|{#1}\right|}
\newcommand{\Error}{\mathrm{Error}}
\newcommand{\zero}{\boldsymbol{0}}
\newcommand{\one}{\boldsymbol{1}}
\newcommand{\limn}{\lim_{n\to\infty}}
\newcommand{\limsupn}{\limsup_{n\to\infty}}
\newcommand{\im}{\mathrm{Im}}
\newcommand{\bcA}{\boldsymbol{\mathcal{A}}}
\newcommand{\bcB}{\boldsymbol{\mathcal{B}}}
\newcommand{\bcP}{\boldsymbol{\mathcal{P}}}
\newcommand{\bpA}{\bp_{\sfA}}
\newcommand{\tbpA}{\widetilde{\bp}_{\sfA}}
\newcommand{\tpA}{\widetilde{p}_{\sfA}}
\newcommand{\bpB}{\bp_{\sfB}}
\newcommand{\pA}{p_{\sfA}}
\newcommand{\pB}{p_{\sfB}}
\newcommand{\alphaA}{\alpha_{\sfA}}
\newcommand{\betaA}{\beta_{\sfA}}
\newcommand{\alphaB}{\alpha_{\sfB}}
\newcommand{\betaB}{\beta_{\sfB}}
\newcommand{\aalphaA}{\aalpha_{\sfA}}
\newcommand{\bbetaA}{\bbeta_{\sfA}}
\newcommand{\aalphaB}{\aalpha_{\sfB}}
\newcommand{\bbetaB}{\bbeta_{\sfB}}
\newcommand{\taalphaA}{\widetilde{\aalpha}_{\sfA}}
\newcommand{\talphaA}{\widetilde{\alpha}_{\sfA}}
\newcommand{\oH}{\overline{H}}
\newcommand{\uH}{\underline{H}}
\newcommand{\uI}{\underline{I}}
\newcommand{\oT}{\overline{\mathcal{T}}}
\newcommand{\uT}{\underline{\mathcal{T}}}

\allowdisplaybreaks

\title{Construction of a Channel Code\\
 from an Arbitrary Source Code\\
 with Decoder Side Information}
\author{
  Jun~Muramatsu
  and~Shigeki Miyake
  \thanks{J.~Muramatsu is with
   NTT Communication Science Laboratories, NTT Corporation,
   2-4, Hikaridai, Seika-cho, Soraku-gun, Kyoto 619-0237, Japan
   (E-mail: muramatsu.jun@lab.ntt.co.jp).
   S.~Miyake is with
   NTT Network Innovation Laboratories, NTT Corporation,
   Hikarinooka 1-1, Yokosuka-shi, Kanagawa 239-0847, Japan
   (E-mail: miyake.shigeki@lab.ntt.co.jp).
   This paper was presented in part at the 2016 International
   Symposium on Information Thoeory and its Applications (ISITA2016),
   Monterey, USA, Oct.~30--Nov.~2, 2016.
  }
}
  
\begin{document}
\maketitle

\begin{abstract}
 The construction of a channel code
 by using a source code with decoder side information
 is introduced.
 For the construction, 
 any pair of encoder and decoder
 is available for a source code with decoder side
 information.
 A constrained-random-number generator,
 which generates random numbers satisfying a condition
 specified by a function and its value,
 is used to construct a stochastic channel encoder.
 The result suggests that
 we can divide the channel coding problem
 into the problems of channel encoding and 
 source decoding with side information.
\end{abstract}

\maketitle

\section{Introduction}

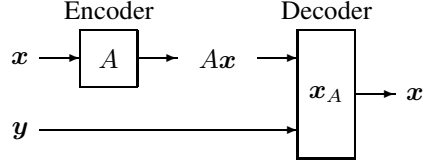
\begin{figure}[h]
\begin{center}
\unitlength 0.53mm
\begin{picture}(105,40)(0,0)
\put(5,25){\makebox(0,0){$\xx$}}
\put(10,25){\vector(1,0){10}}
\put(27,37){\makebox(0,0){Encoder}}
\put(20,18){\framebox(14,14){$A$}}
\put(34,25){\vector(1,0){10}}
\put(54,25){\makebox(0,0){$A\xx$}}
\put(64,25){\vector(1,0){10}}
\put(5,7){\makebox(0,0){$\yy$}}
\put(10,7){\vector(1,0){64}}
\put(81,37){\makebox(0,0){Decoder}}
\put(74,0){\framebox(14,32){$\xx_A$}}
\put(88,16){\vector(1,0){10}}
\put(103,16){\makebox(0,0){$\xx$}}
\end{picture}
\end{center}
\caption{Source Coding with Decoder Side Information:
  An encoder sends a codeword $A\xx$ obtained from a source output $\xx$
  by using an encoding function $A$.
  A decoder reproduces $\xx$ from
  the codeword and side information $\yy$
  by using a decoding function $\xx_A$.}
\label{fig:si}
\end{figure}

The source coding with decoder side information (Fig.~\ref{fig:si})
is a special case of the distributed coding of correlated sources
introduced by Slepian and Wolf~\cite{SW73}.
Let $(X^n,Y^n)$ be a pair of correlated sources.
We consider a source code where
an encoder transmits a codeword obtained from a source output $X^n$
and a decoder reproduces $X^n$
from the codeword and the side information $Y^n$,
where it is expected that the decoding error probability is close to
zero.
From the Slepian-Wolf theorem~\cite{SW73},
the asymptotically optimum encoding rate for stationary memoryless sources
is given by the conditional entropy $H(X|Y)$.
The result is extended to general correlated sources $(\XX,\YY)$ in
\cite{MK95}\cite{SV93}, where conditions such as stationarity and ergodicity are
not assumed and the fundamental limit is given by the
conditional spectral sup-entropy rate $\oH(\XX|\YY)$.
This paper considers a pair of general correlated sources $(\XX,\YY)$,
and the results can be applied to the stationary memoryless case.

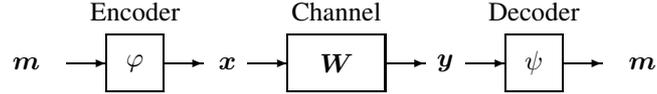
\begin{figure}[h]
\begin{center}
\unitlength 0.53mm
\begin{picture}(153,23)(0,0)
\put(0,7){\makebox(0,0){$\mm$}}
\put(10,7){\vector(1,0){10}}
\put(27,20){\makebox(0,0){Encoder}}
\put(20,0){\framebox(14,14){$\vphi$}}
\put(34,7){\vector(1,0){10}}
\put(50,7){\makebox(0,0){$\xx$}}
\put(55,7){\vector(1,0){10}}
\put(77,20){\makebox(0,0){Channel}}
\put(65,0){\framebox(24,14){$\WW$}}
\put(89,7){\vector(1,0){10}}
\put(104,7){\makebox(0,0){$\yy$}}
\put(109,7){\vector(1,0){10}}
\put(126,20){\makebox(0,0){Decoder}}
\put(119,0){\framebox(14,14){$\psi$}}
\put(133,7){\vector(1,0){10}}
\put(153,7){\makebox(0,0){$\mm$}}
\end{picture}
\end{center}
\caption{Channel Coding:
An encoder sends a codeword $\xx$ obtained from a message $\mm$
by using an encoding function $\vphi$.
A decoder receives an output $\yy$ of a channel $\WW$
and reproduces $\mm$ from $\yy$ by using a decoding function $\psi$.}
\label{fig:channel}
\end{figure}

The main result of this paper is that
we can construct a channel code (Fig.~\ref{fig:channel}) 
from a given source code of $\XX$ with
decoder side information $\YY$,
where the channel input and output are given by
$\XX$ and $\YY$, respectively.
We can construct a code that achieves the capacity
by letting $\XX$ be an optimum channel input random variable
and using a source code achieving the limit $\oH(\XX|\YY)$.

Historically, the Slepian-Wolf codes are constructed by using channel codes.
In~\cite{SW73},
the code is given by using a set of randomly-generated channel codewords
that covers the conditionally typical set of $X$ for a given $Y$.
Wyner~\cite{W74} introduced the Slepian-Wolf code
by using parity check matrices,
where it is shown by Elias~\cite{E55} that
the capacity of a binary symmetric channel
is achievable by using a linear or a convolutional code.
In accordance with this idea,
the Slepian-Wolf codes are constructed
in~\cite{BM01,GZ01,HKU09,M02}
from turbo codes~\cite{BGT93},
polar codes~\cite{A09},
and low density parity check (LDPC) codes~\cite{GA62}
with practical decoding algorithms.
It should be noted that the correlation of two sources are assumed to be
binary-symmetric.
Codes for an asymmetric channel
can be constructed by using
the channel-input alphabet extention ~\cite{BB04}, \cite[Sec.~6.2]{GA68}
or polar codes~\cite{HY13}.

On the other hand, Cover~\cite{C75}
introduced the random binning method for constructing
the Slepian-Wolf code,
where conditions such as symmetric correlations are not assumed for two
sources.
Following this idea, Csisz\'{a}r~\cite{CSI82} proved
that the fundamental limit is achievable by using a linear code.
In~\cite{SWLDPC},
it is proved that the fundamental limit is achievable by using an LDPC
code.
These results are unified by
using the $2$-universal class of hash functions~\cite{CW} 
and its extensions~\cite{HASH,HASH-BC}.
It should be noted that the use of a typical-set decoder or
a maximum-likelihood decoder is assumed in these results.

Based on the concept of hash functions,
in this paper we adopt an approach
where we construct a channel code from a source code with decoder side
information,
which is a kind of Slepian-Wolf code.
A similar approach is investigated
in the context of the linear codebook-level duality
of channel codes and the Slepian-Wolf codes~\cite{CHJLY09},
where the symmetric correlation of two sources
(channel input and output) is assumed.
This paper does not assume such correlations.
It should be noted that this approach is investigated
in~\cite{MM08,CRNG,HASH,YAG12},
where these papers prove that
there is a pair consisting of a source code with decoder side information
and a encoding map to construct a channel code.
However, a maximal-likelihood decoder is assumed,
and it is unknown whether
for an arbitrary given source code with decoder side information
there is a good encoding map with which to construct a channel code.
In~\cite{SWLDPC},
it is proved by assuming a stationary memoryless condition that
for a given arbitrary linear source code with decoder side information
there is a good encoding map with which to construct a channel code,
where the encoding map is intractable.
In contrast, this paper introduces a tractable encoding map
by using a constrained-random-number generator~\cite{CRNG}.
We can use {\em any} source code with decoder side information,
where it is confirmed theoretically or empirically
that the decoding error probability is small.
Neither a typical-set decoder nor a maximal-likelihood decoder
is assumed for the source code with decoder side information.
Our result suggests that
we can divide channel coding problem
into the problems of channel encoding and source decoding
with side information.
It should be noted that the similar results have been appeared
in~\cite[Remark 2]{RR11}, \cite{YAG12} when the output distribution of the
encoder with side information is close to a uniform disbribution.
In contrast, this paper clarifies that such an assumption is
unnecessary.

This paper is organized as follows.
In Section~\ref{sec:capacity}, we review the formula of the channel capacity.
In Section~\ref{sec:channel},
we introduce the construction of a channel code
by using an arbitrary source code with decoder side information.
Based on the results in Section~\ref{sec:channel},
Section~\ref{sec:crng} revisits the channel code
using the constrained-random-number generator
introduced in~\cite{CRNG}.

\section{Channel Capacity}
\label{sec:capacity}

This section reviews the definition of the capacity
of a general channel.
All the results in this paper are presented by using the
information spectrum method introduced
in~\cite{HV93,HAN,VH94},
where the consistency and stationarity are not assumed.

Let $\X$ and $\Y$ be the alphabets of a channel input and 
output, respectively.
Then product sets $\X^n$ and $\Y^n$ are the alphabets of a channel
input vector $X^n$ and a channel output vector $Y^n$, respectively.
It should be noted that $\X$ and $\Y$ are allowed to be
infinite/uncountable/continuous sets on condition
that probability distributions/measures $\mu_{X^n}$ and
$\mu_{Y^n|X^n}(\cdot|\xx)$, $\xx\in\X^n$
are well-defined.

We consider a general source and a general channel.
A general source $\XX$ is
defined by a sequence $\XX\equiv\{\mu_{X^n}\}_{n=1}^{\infty}$
of probability distributions
and a general channel is defined by
a sequence $\WW\equiv\{\mu_{Y^n|X^n}\}_{n=1}^{\infty}$
of conditional probability distributions.

Here, we define the operational channel capacity
with a channel input constraint specified by
a set $\cP_n$ of probability distributions on $\X^n$.
A typical example of a channel input constraint
is the cost constraint, where
any distribution $\mu\in\cP_n$ satisfies
\[
  \int c_n(\xx)\mu(\xx)d\xx < C
\]
for a given cost function $c_n:\X^n\to[0,\infty)$ and $C\in[0,\infty)$.
\begin{df}
Let $\bcP\equiv\{\cP_n\}_{n=0}^{\infty}$ be a sequence
of the set of probability distributions on $\X^n$.
For a general channel $\WW$,
we call a rate $R$ {\em achievable} if for all
$\delta>0$
and all sufficiently large $n$ there is a pair consisting of
an encoder $\vphi_n:\M_n\to\cS^n$
and a decoder $\psi_n:\Y^n\to\M_n$ such that
\begin{gather}
  \frac 1n\log|\M_n|\geq R
  \label{eq:capacity-rate}
  \\
  \mu_{X^n}\in\cP_n
  \label{eq:capacity-constraint}
  \\
  P(\psi_n(Y^n)\neq M_n)\leq\delta,
  \label{eq:capacity-error}
\end{gather}
where
we call a subset $\cS$ of $\X$ a {\em signaling alphabet}\footnote{This
 terminology comes from~\cite{B87}.},
$\M_n$ is a set of messages,
$[1/n]\log|\M_n|$ represents the rate of the code,
$M_n$ is a random variable of the message
corresponding to the uniform distribution on
$\M_n$,
$Y^n$ is the random variable of a channel output
with an input $X^n\equiv\vphi_n(M_n)$,
and the joint distribution $\mu_{M_nY^n}$ is given as
\[
 \mu_{M_nY^n}(\mm,\yy)\equiv\frac{\mu_{Y^n|X^n}(\yy|\vphi_n(\mm))}{|\M_n|}.
\]
The {\em channel capacity} $C_{\cS}(\WW)$ is defined by the supremum
of the achievable rate, where the signaling alphabet $\cS$ is specified.
\end{df}

It should be noted that
the standard definition of channel capacity
can be denoted by $C_{\X}(\WW)$,
where the signaling alphabet $\cS$ is equal to $\X$.
It should also be noted that
we can let $\cP^n$ be the set of all probability distributions on $\X^n$
when it is assumed that there is no channel input constraint.

Next, let us define the capacity $C_{\X}^q(\WW)$
of a channel with a finite signaling alphabet as
\[
  C_{\X}^q(\WW)
  \equiv
  \sup_{\substack{
      \cS\subset\X:
      |\cS|\leq q
  }}
  C_{\cS}(\WW).
\]
Since $\{\cS: |\cS|\leq q\}\subset\{\cS:|\cS|\leq q+1\}$,
we have the fact that 
$C_{\X}^q(\WW)$ is a non-decreasing function of $q$.
We have the following lemma.
\begin{lem}[{\cite[Theorem 2]{COUNTABLELIMIT}}]
\label{lem:countable}
When $C_{\X}(\WW)<\infty$, we have
\[
  C_{\X}(\WW)=\lim_{q\to\infty}C_{\X}^q(\WW).
\]
\end{lem}

The proof is given in Section~\ref{sec:proof-countable}
for the completeness of this paper.

For a general channel $\WW$, the channel capacity $C_{\X}(\WW)$
is derived in \cite{VH94}, \cite[Theorem~3.6.1]{HAN}\footnote{
In~\cite[Theorem~3.6.1]{HAN}, it is assumed that $\bcP$
is a cost constraint. However, we can easily extend
the result to an arbitrary channel input constraint.}
as
\begin{equation}
  C_{\X}(\WW)=\sup_{\XX\in\bcP}\uI(\XX;\YY),
  \label{eq:capacity-I}
\end{equation}
where the supremum is taken over all general sources
$\XX=\{\mu_{X^n}\}_{n=1}^{\infty}$
such that $\mu_{X^n}\in\cP_n$ for every $n$,
and the joint distribution $\mu_{X^nY^n}$ is given as
\begin{equation}
  \mu_{X^nY^n}(\xx,\yy)
  \equiv
  \mu_{Y^n|X^n}(\yy|\xx)\mu_{X^n}(\xx).
  \label{eq:channel-joint}
\end{equation}
Furthermore, similarly to the proof in~\cite{CRNG},
we can show the formula
\begin{equation}
  C_{\X}(\WW)=\sup_{\XX\in\bcP}\lrB{\uH(\XX)-\oH(\XX|\YY)}
  \label{eq:capacity}
\end{equation}
when $\X$ is finite, where the supremum is taken over all general sources
$\XX$ and the joint distribution of $(\XX,\YY)$ is given
by (\ref{eq:channel-joint}).
From Lemma \ref{lem:countable},
we have the following lemma.

\begin{lem}
When $C_{\X}(\WW)<\infty$, we have
\begin{align*}
  C_{\X}(\WW)
  &=
  \lim_{q\to\infty}
  \sup_{\substack{
      \cS\subset\X:\\
      |\cS|\leq q
  }}
  \sup_{\substack{
      \XX\in\bcP:
      \\
      X^n\in\cS^n\ \text{for all}\ n
  }}
  \uI(\XX;\YY)
  \\
  &=
  \lim_{q\to\infty}
  \sup_{\substack{
      \cS\subset\X:\\
      |\cS|\leq q
  }}
  \sup_{\substack{
      \XX\in\bcP:
      \\
      X^n\in\cS^n\ \text{for all}\ n
  }}
  [\uH(\XX)-\oH(\XX|\YY)],
\end{align*} 
where the condition $X^n\in\cS^n$ implies
that the support of the probability distribution
of a channel input is a subset of $\cS^n$.
\end{lem}

\begin{rem}
For many channels it is known that
an optimal input distribution in a channel coding
has a discrete support, where a support is defined as the set
of all elements with positive measure.
For example, for an additive white Gaussian noise (AWGN) channel,
it is shown in~\cite{S71} that
the optimal input distribution has a discrete and finite support
under the maximum power constraint.
It should be noted that the above lemma implies that
we can approach the capacity with a sufficiently large
signaling alphabet for any channel with an uncountable/continuous
channel input alphabet (e.g.\ AWGN channel under the avarage power constraint).
\end{rem}

In this paper, we show the fact that
for a given finite set $\X$
there is a code such that the rate of the code is close to the right
hand side of (\ref{eq:capacity}).
Then, from the above lemma,
we have the fact that
the capacity of a channel with an uncountable channel input alphabet
is achievable with the code
by optimizing the finite signaling alphabet $\cS$ and letting $|\cS|\to\infty$.

\begin{rem}
In~\cite[Section~7.8]{B87}, the optimal signaling alphabet $\cS\in\X$
is derived for an additive white Gaussian noise channel, where it is
assumed that
all the symbols in $\cS$ are used equally often, that is, the input
distribution is uniform on $\cS$.
This assumption is natural when we use conventional linear codes.
On the other hand,
it is unnecessary to assume that the input distribution is uniform on $\cS$
in the code construction introduced in~\cite{CRNG},
where the encoding rate may increase.
\end{rem}

\section{Channel Code by Using Source Code with Decoder Side Information}
\label{sec:channel}

In this section, we construct a channel code
by using a source code with decoder side information.
To this end, we review a balanced-coloring property~\cite{CRNG-VLOSSY},
which is a variant of the hash property
introduced in~\cite{CRNG,HASH,ISIT2010,ISIT2011a}.

\subsection{$(\aalpha,\bbeta)$-Balanced-Coloring Property}
\label{sec:hash}

Throughout this paper, we use the following definitions and notations.
The complement of $\U$ is denoted by $\U^c$
and the set difference is defined as $\U\setminus\V\equiv\U\cap\V^c$.
Let $B\xx$ denote a value taken by a function $B$ at $\xx\in\X^n$,
where $B$ may be nonlinear.
When $B$ is a linear function
expressed by an $l\times n$ matrix,
we assume that $\X\equiv\GFq$
is a finite field and the range of functions is $\X^l$.
For a function $B$ and a set $\B$ of functions, 
let $\im B$ and $\im\B$ be defined as
\begin{align*}
  \im B&\equiv\{B\xx: \xx\in\X^n\}
  \\
  \im\B &\equiv \bigcup_{B\in\B}\im B.
\end{align*}
We define a set $\C_B(\mm)$ as
\begin{align*}
  \C_B(\mm) &\equiv\{\xx: B\xx = \mm\}.
\end{align*}
The random variables of a function $B$ and a vector $\cc$ are denoted
by the sans serif letters $\sfB$ and $\sfcc$, respectively.
It should be noted that
the random variable of
an $n$-dimensional vector $\xx\in\X^n$ is denoted by
the Roman letter $X^n$ that does not represent a function,
which is the way it has been used so far.
The symbol $E$ denotes the expectation.
For example, $E_{\sfB\sfcc}[\cdot]$ denotes the expectation
with respect to random variables $\sfB$ and $\sfcc$.

Here, we introduce the balanced-coloring property for an
ensemble of functions.
It requires weaker conditions than
the hash property introduced in~\cite{CRNG,ISIT2010,ISIT2011a}.
\begin{df}[\cite{CRNG-VLOSSY}]
Let $\B_n$ be a set of functions on $\X^n$
and $p_{\sfB,n}$ be a probability distribution on $\B_n$.
We call a pair $(\B_n,p_{\sfB,n})$ an {\em ensemble}.
Then a sequence $(\bcB,\bpB)\equiv\{(\B_n,p_{\sfB,n})\}_{n=1}^{\infty}$
has an {\em $(\aalphaB,\bbetaB)$-balanced-coloring property} if
there are two sequences
$\aalphaB{}\equiv\{\alphaB(n)\}_{n=1}^{\infty}$ and
$\bbetaB{}\equiv\{\betaB(n)\}_{n=1}^{\infty}$,
depending on $\{p_{\sfB,n}\}_{n=1}^{\infty}$,
such that
\begin{align}
  &\limsupn \alphaB(n)= 1
  \tag{BC1}
  \label{eq:alpha}
  \\
  &\limsupn\frac1n\log(\betaB(n)+1)= 0
  \tag{BC2}
  \label{eq:beta}
\end{align}
and
\begin{align}
  \sum_{\substack{
      \xx'\in\X^n\setminus\{\xx\}:
      \\
      p_{\sfB,n}(\{B: B\xx = B\xx'\})>\frac{\alphaB(n)}{|\im\B_n|}
  }}
  p_{\sfB,n}\lrsb{\lrb{B: B\xx = B\xx'}}
  \leq
  \betaB(n)
 \tag{BC3}
 \label{eq:hash}
\end{align}
for all sufficiently large\footnote{In~\cite{CRNG-VLOSSY},
an ensemble is required to satisfy (\ref{eq:hash}) for all $n$
and all $\xx\in\X^n$. However, it is sufficient to assume that
an ensemble satisfies (\ref{eq:hash})
for sufficiently large $n$ and all $\xx\in\X^n$
because we finally let $n\to\infty$.}
$n$ and all $\xx\in\X^n$.
Throughout this paper, we
omit the dependence of
$\B$, $\pB$, $\alphaB$ and $\betaB$ on $n$.
\end{df}

Here, let us introduce examples
satisfying the balanced-coloring property.
When $\B$ is a two-universal class of hash functions \cite{CW}
and  $\pB$ is the uniform distribution on $\B$,
then $(\bcB,\bpB)$ has a $(\one,\zero)$-balanced-coloring property,
where $\one$ and $\zero$ denote the constant sequences 
of $1$ and $0$, respectively.
Random binning~\cite{C75}
and a set of all linear functions \cite{CSI82} are examples of
the two-universal class of hash functions.
It is proved in \cite{CRNG-VLOSSY} that
an ensemble of systematic sparse matrices has a
balanced-coloring property,
where a matrix has an identity sub-matrix with the same number of rows.

The following lemma
is an extension of the leftover hash lemma~\cite{IZ89},
the balanced-coloring
lemma~\cite[Lemma 3.1]{AC98}, \cite[Lemma 17.3]{CK11},
and the output statistics of random binning~\cite{YAG12}.
This lemma implies that there is a function $B$ such that
$\T$ is almost equally partitioned by $B$ with respect to a measure $Q$.
\begin{lem}[{\cite[Lemma 5]{CRNG},\cite[Lemma 4]{ISIT2011a}}]
\label{lem:BCP}
If $(\B,p_{\sfB})$ satisfies (\ref{eq:hash}), then
\begin{align*}
 E_{\sfB}\lrB{
  \sum_{\mm}
  \left|
   \frac{Q\lrsb{\T\cap\C_{\sfB}(\mm)}}{Q(\T)}
   -\frac 1{|\im\B|}
  \right|
 }
 &\leq
 \sqrt{
  \alphaB-1
  +\frac {[\betaB+1]|\im\B|\max_{\uu\in\T} Q(\uu)}{Q(\T)}
 }
\end{align*}
for any function $Q:\X^n\to[0,\infty)$ and $\T\subset\X^n$,
where
\[
 Q(\T)\equiv\sum_{\xx\in\T}Q(\xx).
\]
\end{lem}

\subsection{Construction of Channel Code}
This section introduces a channel code.
The idea for the construction is drawn
from~\cite{HASH,ISIT2011a,SWLDPC}.
We assume that
the channel input alphabet $\X^n$ is a finite set
but allow the channel output alphabet $\Y^n$ to be an arbitrary
(infinite, continuous) set.

We assume that the channel distribution $\mu_{Y^n|X^n}$
and the input distribution $\mu_{X^n}$ are given.
Let  $\mu_{X^n|Y^n}$ be defined as
\[
  \mu_{X^n|Y^n}(\xx|\yy)\equiv
  \frac{\mu_{Y^n|X^n}(\yy|\xx)\mu_{X^n}(\xx)}
  {\sum_{\xx}\mu_{Y^n|X^n}(\yy|\xx)\mu_{X^n}(\xx)}.
\]

Here, we assume that an arbitrary fixed-length source code
$(A,\xx_A)$
with decoder side information
(Fig.~\ref{fig:si})
is given, where
$A:\X^n\to\im A$ is a encoding function
and $\xx_A:\im A\times\Y^n\to\X^n$ is a decoding function.
Then the coding rate $r$ of the code is given as
\begin{equation}
  r\equiv\frac1n\log|\im A|.
  \label{eq:r}
\end{equation}
The decoding error probability $\Error(A)$ of this code
is given as
\begin{equation}
  \Error(A)\equiv\mu_{X^nY^n}\lrsb{\lrb{
      (\xx,\yy): \xx_A(A\xx|\yy)\neq \xx
  }}.
  \label{eq:sw}
\end{equation}
It should be noted that
the condition $\limn\Error(A)=0$ is not assumed for this code,
and this code may be sub-optimal
in the sense that the coding rate $r$ is not close
to the fundamental limit
described as the conditional spectral sup-entropy rate
$\oH(\XX|\YY)$.

For a given $R>0$, let $(\B,\pB)$ be an ensemble of functions on the
set $\X^n$ satisfying
\begin{equation}
  R=\frac1n\log|\im\B|,
  \label{eq:R}
\end{equation}
where we define $\M_n\equiv\im\B$
and $R$ represents the rate of the code.
We obtain a function $B\in\B$ and a vector $\cc\in\im A$ generated at
random subject to the distribution $\pB$ and
$\{\mu_{X^n}(\C_A(\cc))\}_{\cc\in\im A}$, respectively.
It should be noted that
we can obtain $\cc\equiv A\xx$
generated at random subject to the distribution
$\{\mu_{X^n}(\C_A(\cc))\}_{\cc\in\im A}$
by generating $\xx$ at random subject to the distribution $\mu_{X^n}$
and operating $A$ on  $\xx$.

We fix $B$ and $\cc$ so that they are shared by the channel
encoder and the channel decoder.
To summarize, the channel encoder has
functions $A$, $B$ and a vector $\cc$,
and the channel decoder has functions $\xx_A$, $B$, and a vector $\cc$.

We use a constrained-random-number generator to construct a
stochastic encoder.
Let $\tX^n\equiv\tX^n_{AB}(\cc,\mm)$ be a random variable corresponding
to the distribution
\[
  \nu_{\tX^n|M_n}(\xx|\mm)
  \equiv
  \begin{cases}
    \frac{\mu_{X^n}(\xx)}{\mu_{X^n}(\C_{AB}(\cc,\mm))},
    &\text{if}\ \xx\in\C_{AB}(\cc,\mm),
    \\
    0,
    &\text{if}\ \xx\notin\C_{AB}(\cc,\mm),
  \end{cases}
\]
where  $\C_{AB}(\cc,\mm)\equiv\C_A(\cc)\cap\C_B(\mm)$.
The encoder generates $\xx$ that satisfies $A\xx=\cc$ and $B\xx=\mm$
with probability $\nu_{\tX^n|M_n}(\xx|\mm)$.
It should be noted that we can use
the sum-product algorithm or the Markov-Chain-Monte-Carlo method
to implement the constrained-random-number generator~\cite{CRNG,CRNG-VLOSSY}.
We define the stochastic channel encoder
$\Phi_n:\im\B\to\X^n$
as
\[
  \Phi_n(\mm)
  \equiv
  \begin{cases}
    \tX^n_{AB}(\cc,\mm),
    &\text{if}\ \mu_{X^n}(\C_{AB}(\cc,\mm))>0,
    \\
    \text{``error,''}
    &\text{if}\ \mu_{X^n}(\C_{AB}(\cc,\mm))=0.
  \end{cases}
\]

\begin{figure}[t]
\begin{center}
\unitlength 0.45mm
\begin{picture}(176,70)(0,0)
\put(82,60){\makebox(0,0){Encoder}}
\put(65,35){\makebox(0,0){$\cc$}}
\put(70,35){\vector(1,0){10}}
\put(35,17){\makebox(0,0){$\mm$}}
\put(45,17){\vector(1,0){35}}
\put(80,10){\framebox(18,32){$\tX^n_{AB}$}}
\put(98,26){\vector(1,0){20}}
\put(125,26){\makebox(0,0){$\xx$}}
\put(55,0){\framebox(54,52){}}
\end{picture}
\\
\begin{picture}(176,70)(0,0)
\put(82,60){\makebox(0,0){Decoder}}
\put(30,35){\makebox(0,0){$\cc$}}
\put(35,35){\vector(1,0){10}}
\put(5,17){\makebox(0,0){$\yy$}}
\put(10,17){\vector(1,0){35}}
\put(45,10){\framebox(18,32){$\xx_A$}}
\put(63,26){\vector(1,0){10}}
\put(83,26){\makebox(0,0){$\xx$}}
\put(93,26){\vector(1,0){10}}
\put(103,19){\framebox(30,14){$B$}}
\put(133,26){\vector(1,0){20}}
\put(162,26){\makebox(0,0){$\mm$}}
\put(20,0){\framebox(124,52){}}
\end{picture}
\end{center}
\caption{Construction of Channel Code}
\label{fig:channel-code}
\end{figure}
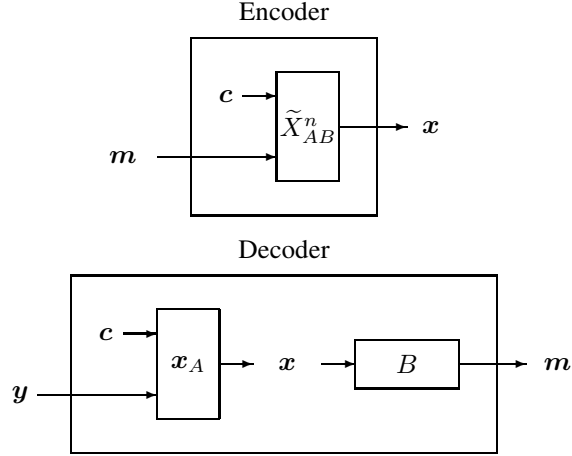

Let $\yy\in\Y^n$ be a channel output.
We define the channel decoder $\psi_n:\Y^n\to\im\B$ as
\[
  \psi_n(\yy)
  \equiv B\xx_A(\cc|\yy),
\]
where the decoder reproduces $\xx$ that satisfies $A\xx=\cc$
by using $\xx_A$
and obtains a reproduced message $\mm=B\xx$.
The flow of vectors is illustrated in Fig.~\ref{fig:channel-code}.

The error probability $\Error(A,B,\cc)$ is given by
\[
 \begin{split}
  \Error(A,B,\cc)
  &\equiv
  \sum_{\substack{
    \mm:
    \mu_{X^n}(\C_{AB}(\cc,\mm))=0
  }}
  \frac1
  {|\M_n|}
  +
  \sum_{\substack{
    \mm,\xx,\yy:\\
    \mu_{X^n}(\C_{AB}(\cc,\mm))>0\\
    \xx\in\C_{AB}(\cc,\mm)\\
    \psi_n(\yy)\neq\mm
  }}
  \frac{\mu_{Y^n|X^n}(\yy|\xx)\mu_{X^n}(\xx)}
  {|\M_n|\mu_{X^n}(\C_{AB}(\cc,\mm))}.
 \end{split}
 \label{eq:channel-error-stochastic}
\]
We have the following theorem, where the proof is given in
Section~\ref{sec:proof-channel}.
\begin{thm}
\label{thm:channel}
Let $(A,\xx_A)$ be a source code with decoder side information,
where the encoding rate and the decoding error probability
are given by (\ref{eq:r}) and (\ref{eq:sw}), respectively.
Assume that $(\bcB,\bpB)$ 
has an $(\aalphaB,\bbetaB)$-balanced-coloring property
for a given $R$ satisfying (\ref{eq:R}) and
\begin{align}
  r+R&<\uH(\XX),
  \label{eq:channel-rR}
\end{align}
where $\uH(\XX)$ is the spectral inf-entropy rate.
Then for any $\delta>0$
and all sufficiently large $n$
there is a function $B\in\B$, and a vector $\cc\in\im\A$ such that
\begin{equation}
  \Error(A,B,\cc)\leq\Error(A)+\delta.
  \label{eq:channel-error}
\end{equation}
\end{thm}

\begin{rem}
In~\cite[Remark 2]{RR11}, \cite{YAG12},
inequality (\ref{eq:channel-error})
is shown by assuming that the output distribution of
an encoding function $A$ is close to a uniform distribution.
In contrast, such an assumption is not assumed in the above theorem.
\end{rem}

From the above theorem and (\ref{eq:capacity}), the channel capacity
is achievable with the proposed code
by letting $\XX$ be a source that attains the supremum
on the right hand side of (\ref{eq:capacity}),
$R\to\uH(\XX)-\oH(\XX|\YY)$, $\delta\to0$,
and assuming that $(A,\xx_A)$ is optimum
in the sense that $r\to\oH(\XX|\YY)$ and $\Error(A)\to0$.

\section{Channel Code Using Constrained-Random-Number Generator
  Revisited}
\label{sec:crng}
In this section, we revisit the channel code
using the constrained-random-number generator
introduced in~\cite{CRNG}.
To this end, we introduce a variant of the hash
property~\cite{CRNG,HASH}
and a construction of source code with decoder side information
based on this property.

\subsection{$(\aalpha,\bbeta)$-Collision-Resistance Property}
\label{sec:cr}

In this section, we revisit \cite[Remark 1]{HASH}, which mentions
that some ensembles of sparse matrices
satisfy the weaker condition $\limn[1/n]\log\alpha_A(n)=0$.
We introduce the collision-resistance property as follows.

\begin{df}
Let $\A_n$ be a set of functions on $\X^n$
and $p_{\sfA,n}$ be a probability distribution on $\A_n$.
Then a sequence $(\bcA,\bpA)\equiv\{(\A_n,p_{\sfA,n})\}_{n=1}^{\infty}$
has an {\em $(\aalphaA,\bbetaA{})$-collision-resistance property} if
there are two sequences
$\aalphaA{}\equiv\{\alphaA(n)\}_{n=1}^{\infty}$ and
$\bbetaA{}\equiv\{\betaA(n)\}_{n=1}^{\infty}$,
depending on $\{p_{\sfA,n}\}_{n=1}^{\infty}$,
such that
\begin{align}
  &\limsupn \frac1n\log\alphaA(n)= 0
  \tag{CR1}
  \label{eq:alpha-cr}
  \\
  &\limsupn\betaA(n)= 0
  \tag{CR2}
  \label{eq:beta-cr}
\end{align}
and
\begin{align}
  \sum_{\substack{
      \xx'\in\X^n\setminus\{\xx\}:
      \\
      p_{\sfA,n}(\{A: A\xx = A\xx'\})>\frac{\alphaA(n)}{|\im\A_n|}
  }}
  p_{\sfA,n}\lrsb{\lrb{A: A\xx = A\xx'}}
  \leq
  \betaA(n)
 \tag{CR3}
 \label{eq:hash-cr}
\end{align}
for all sufficiently large $n$ and all $\xx\in\X^n$,
where (\ref{eq:hash-cr}) is the same as (\ref{eq:hash}).
\end{df}

It should be noted that
when $\A$ is a two-universal class of hash functions~\cite{CW}
and  $\pA$ is the uniform distribution on $\A$,
then $(\bcA,\bpA)$ also has a $(\one,\zero)$-collision-resistance property.
Random binning~\cite{C75}
and a set of all linear functions \cite{CSI82} are examples of
the two-universal class of hash functions.
The motivation for introducing the collision-resistance property
is that we can include the expurgated ensemble of linear functions
introduced in the next section.

The following lemma is related to the
{\it collision-resistance property}, that is,
if the number of bins is greater than the number of items
then there is an assignment such that every bin contains at most one item.
\begin{lem}[{\cite[Lemma 4]{CRNG},\cite[Lemma 1]{HASH}}]
\label{lem:CRP}
If $(\A,\pA)$ satisfies (\ref{eq:hash-cr}),
then
\[
  \pA\lrsb{\lrb{
      A: \lrB{\G\setminus\{\uu\}}\cap\C_A(A\uu)\neq \emptyset
  }}
  \leq 
  \frac{|\G|\alphaA{}}{|\im\A|} + \betaA{}
\]
for all $\G\subset\U^n$  and $\uu\in\U^n$.
\end{lem}

\subsection{Expurgated Ensemble of Linear Functions}

In this section, we consider the expurgated ensemble of linear
functions. The idea comes from~\cite{BB04,EM05,GA62,MB01}.

Let $\X$ be a finite field
and  $\A$ be a set of
linear functions $A:\X^n\to\X^l$,
where $A$ can be expressed by an $l\times n$ matrix.
Let $\bt(\xx)$ be the type
of  $\xx\in\U^n$,
which is characterized by the empirical probability distribution
of the sequence $\xx$.
For a type $\bt$, let $\C_{\bt}$ be defined as
\[
  \C_{\bt}\equiv\{\xx\in\X^n: \bt(\xx)=\bt\}.
\]
Let $\cH$ be the set of all types of length $n$ except $\bt(\zero)$,
where $\zero$ is the zero vector.
For a probability distribution
$\pA$ on a set of $l\times n$ matrices
and a type $\bt$, let
$S(\pA,\bt)$ be defined as
\begin{align*}
  S(\pA,\bt)
  &\equiv
  \sum_{A\in\A}\pA(A)|\{\xx\in\X^n: A\xx=\zero, \bt(\xx)=\bt\}|
  \notag
  \\
  &=
  \sum_{A\in\A}\pA(A)|\C_A(\zero)\cap\C_{\bt}|,
\end{align*}
which is the expected number of codewords that have type $\bt$.
For a given $\hcH\subset\cH$, we define $\alphaA(n)$ and
$\betaA(n)$ as
\begin{align}
 \alphaA(n)
 &\equiv
 \frac{|\im\A|}{|\X|^l}\cdot\max_{\bt\in \hcH}
 \frac {S(\pA,\bt)}{S(p_{\osfA},\bt)}
 \label{eq:alpha-linear}
 \\
 \betaA(n)
 &\equiv
 \sum_{\bt\in \cH\setminus\hcH}S(\pA,\bt),
 \label{eq:beta-linear}
\end{align}
where $p_{\osfA}$ denotes the uniform distribution
on the set of all $l\times n$ matrices.

For an ensemble $(\A,\pA)$
of linear functions,
we define an expurgated ensemble $(\A,\tpA)$
as
\[
  \tpA(A)
  \equiv
  \begin{cases}
    \frac{\pA(A)}
    {\pA\lrsb{\lrb{
	  A: 
	  \bt(\xx)\in\hcH\cup\{\bt(\zero)\}
	  \ \text{for all}\ \xx\in\C_A(\zero)
      }}
    }
    \\
    \qquad\text{if}\ \bt(\xx)\in\hcH\cup\{\bt(\zero)\}
    \ \text{for all}\ \xx\in\C_A(\zero),
    \\
    0
    \hspace{5mm}\text{if}\ \exists \xx\in\C_A(\zero)
    \ \text{s.t.}\ \bt(\xx)\in\cH\setminus\hcH,
  \end{cases}
\]
where $\hcH$ will be specified later.
It should be noted that $\hcH$ is usually defined
as the set of $\bt$ satisfying
the requirement that
the weight (the number of non-zero symbols)
is greater than $\gamma n$, where $\gamma$ depends only on $|\X|$ and
$l$.
Then the condition
$\bt(\xx)\in\hcH\cup\{\bt(\zero)\}$
for all $\xx\in\C_A(\zero)$
implies that
the minimum distance of $A$
(the minimum Hamming distance between two different vectors in $\C_A(\zero)$)
is greater than $\gamma n$.
That is,
the ensemble $(\A,\tpA)$
is obtained from $(\A,\pA)$
by expurgating functions with the minimum distance 
smaller than $\gamma n$.

We have the following lemma,
where the proof is given in Section~\ref{sec:proof-ex}.
\begin{lem}
\label{lem:ex}
We first assume that
for an ensemble $(\A,\pA)$
of linear functions
there is $\hcH$ such that
a pair $\alphaA$ and $\betaA$ defined by (\ref{eq:alpha-linear}) and
(\ref{eq:beta-linear}), respectively,
satisfy
\begin{align}
  &\limsupn \frac1n\log\alphaA(n)=0
  \label{eq:alpha-ex}
  \\
  &\limsupn \frac1n\log\frac1{1-\betaA(n)}=0,
  \label{eq:beta-ex}
\end{align}
where (\ref{eq:beta-ex}) implicitly assumes that
$\betaA<1$ for all sufficiently large $n$.
Then the sequence $(\bcA,\tbpA)$ of expurgated ensembles
satisfies the $(\taalphaA,\zero)$-collision-resistance property,
where $\taalphaA\equiv\{\talphaA(n)\}_{n=1}^{\infty}$
is given as
\[
  \talphaA(n)
  \equiv \frac{\alphaA(n)}{1-\betaA(n)}.
\]
Furthermore, we have
\begin{equation}
  \tpA\lrsb{\lrb{
      A: \lrB{\G\setminus\{\xx\}}\cap\C_A(A\xx)\neq \emptyset
  }}
  \leq 
  \frac{|\G|\alphaA}{|\im\A|[1-\betaA]}
  \label{eq:CRPex}
\end{equation}
for all sufficiently large $n$, all $\G\subset\X^n$, and all $\xx\in\X^n$.
\end{lem}

\subsection{Source Code with Decoder Side Information}
Here, we consider the source coding
with decoder side information
illustrated in Fig.~\ref{fig:si}.
The fundamental limit for this problem is
given as the conditional spectral sup-entropy rate $\oH(\XX|\YY)$.

The achievability of this problem
is proved via the Slepian-Wolf theorem
using random binning~\cite{C75,MK95,SV93}
or the ensemble of all $q$-ary linear
matrices~\cite{CSI82}.
The construction of an encoder using sparse matrix is studied
in \cite{M02,SPR02}
and the achievability is proved
in \cite{HASH-BC,SWLDPC} by using a maximum-likelihood or
minimum-divergence decoding.
We obtain the coding theorem based on the collision-resistance property
as a corollary of \cite[Theorem 7]{HASH} for stationary memoryless sources.

We fix a function $A:\X^n\to\im\A$
which is used as a encoding function.
The encoding rate $r$ is given as
\begin{align}
  r=\frac{\log|\im\A|}n.
  \label{eq:rate-sir}
\end{align}
We define the decoder
$\xx_A:\im\A\times\Y^n\to\X^n$
as
\begin{align}
  \xx_A(\cc|\yy)
  &\equiv
  \arg\max_{\xx\in\C_A(\cc)}
  \mu_{X^nY^n}(\xx,\yy)
  \label{eq:si-ml}
  \\ 
  &
  =
  \arg\max_{\xx\in\C_A(\cc)}
  \mu_{X^n|Y^n}(\xx|\yy).
  \label{eq:si-map}
\end{align}
The decoding error probability $\Error(A)$ is given as
\begin{equation*}
 \Error(A)
 \equiv
 \mu_{X^nY^n}\lrsb{\lrb{
 (\xx,\yy): \xx_A(A\xx|\yy)\neq \xx
 }}.
\end{equation*}
It should be noted that the construction
is analogous to the syndrome encoding/decoding when $A$ is a
linear function.

We have the following theorem.
It should be noted that
$\Y$ is allowed to be an infinite/continuous set
and the correlation of the two sources is allowed to be asymmetric.
\begin{thm}
\label{thm:si}
Let $(\XX,\YY)$ be a pair of correlated general sources.
Assume that an ensemble $(\bcA,\bpA)$
has an $(\aalphaA,\bbetaA)$-collision-resistance property
for a given $r$ satisfying
(\ref{eq:rate-sir}) and
\begin{equation}
  r > \oH(\XX|\YY).
  \label{eq:sir}
\end{equation}
Then for any $\delta>0$ and all sufficiently large $n$
there is a function (sparse matrix) $A\in\A$ such that
\[
  \Error(A)
  \leq
  \delta.
\]
\end{thm}

It should be noted that the theorem is implicitly proved
in~\cite[Eq.~(58)]{CRNG} from Lemma~\ref{lem:CRP}.
The proof is given in Section~\ref{sec:proof-si}
for the completeness of the paper.

\begin{rem}
\label{rem:stochastic}
We can use a stochastic decoder instead of
the decoder defined by (\ref{eq:si-ml}) or (\ref{eq:si-map}) in the above
coding scheme.
Let
$\chi(\cdot)$ be defined as
\begin{equation}
  \chi(\mathrm{S})
  \equiv
  \begin{cases}
    1,&\text{if the statement $\mathrm{S}$ is true}
    \\
    0,&\text{if the statement $\mathrm{S}$ is false}.
  \end{cases}
  \label{eq:chi}
\end{equation}
Then the joint distribution $p_{X^nY^nC_n}$ is given as
\[
 p_{X^nY^nC_n}(\xx,\yy,\cc)
 =
 \mu_{X^nY^n}(\xx,\yy)\chi(A\xx=\cc),
\]
and the reproduction is determined at random
subject to the distribution
\begin{align}
 p_{X^n|Y^nC_n}(\xx|\yy,\cc)
 &\equiv
 \frac{\mu_{X^nY^n}(\xx,\yy)\chi(A\xx=\cc)}
 {\sum_{\xx}\mu_{X^nY^n}(\xx,\yy)\chi(A\xx=\cc)}
 \notag
 \\
 &=
 \frac{\mu_{X^n|Y^n}(\xx|\yy)\chi(A\xx=\cc)}
 {\mu_{X^n|Y^n}(\C_A(\xx)|\yy)}.
 \label{eq:si-crng}
\end{align}
Then, from Lemma~\ref{lem:sdecoding} in Appendix,
the decoding error probability
is at most twice the error probability
of the decoder defined by (\ref{eq:si-ml}) or (\ref{eq:si-map}).
Thus, from Theorem~\ref{thm:si},
we have the fact that the fundamental limit $\oH(\XX|\YY)$
is also achievable with the code by using this stochastic decoder.
It should be noted that we can consider (\ref{eq:si-crng})
as the output distribution of a constrained-random-number
generator~\cite{CRNG},
which is implementable by using
the sum-product algorithm or the Markov-Chain-Monte-Carlo method
when $(X^n,Y^n)$ is memoryless (see \cite{CRNG,CRNG-VLOSSY,SDECODING}).
\end{rem}

\subsection{Channel Code Using Constrained-Random-Number Generator}

From Theorems~\ref{thm:channel} and~\ref{thm:si},
we have the following corollary,
which is an improvement of~\cite[Theorem~1]{CRNG}.
It should be noted that the conditions for $(\bcA,\bpA)$
and $(\bcB,\bpB)$ are weaker than those in~\cite[Theorem~1]{CRNG}.

\begin{cor}
\label{cor:crng}
Assume that $(\bcA,\bpA)$ and $(\bcB,\bpB)$ have
an $(\aalphaA,\bbetaA)$-collision-resistance property
and an $(\aalphaB,\bbetaB)$-balanced-coloring property,
respectively,
for given $r$ and $R$ satisfying
\begin{align*}
  r&=\frac 1n\log|\im\A|
  \\
  R&=\frac 1n\log|\im\B|
  \\
  r&>\oH(\XX|\YY)
  \\
  r+R&<\uH(\XX).
\end{align*}
Then for any $\delta>0$
and all sufficiently large $n$
there are functions $A\in\A$, $B\in\B$, and
a vector $\cc\in\im\A$ such that the decoding error probability
is less than $\delta$.
The channel capacity is achievable with the proposed code
by letting $\XX$ be a source that attains the supremum
on the right hand side of (\ref{eq:capacity}).
\end{cor}

It should be noted that we can use the stochastic decoder
subject to the distribution defined by (\ref{eq:si-crng})
instead of the decoder defined by (\ref{eq:si-ml}) or (\ref{eq:si-map}).
This implies that both encoding and decoding functions
can be constructed by using constrained-random-number
generators\footnote{We would like to call this type of codes
CoCoNuTS (Codes based on Constrained Numbers
Theoretically-achieving the Shannon limit).}.

\section{Proofs}
\label{sec:proof}

In the following proofs,
we omit the dependence on $n$ of $X$ and $Y$ when they appear
in the subscripts of $\mu$, $\oT$, and $\uT$.
The integral over the alphabet $\Y^n$ is
denoted by $\sum$.

\subsection{Proof of Lemma~\ref{lem:countable}}
\label{sec:proof-countable}
The proof is completed by showing that
\[
  C_{\X}(\WW)\leq \lim_{q\to\infty}C_{\X}^q(\WW)
\]
because we can show 
\[
  C_{\X}(\WW)\geq \lim_{q\to\infty}C_{\X}^q(\WW)
\]
from the trivial fact $C_{\X}(\WW)\geq C_{\X}^q(\WW)$.

From the definition of $C_{\X}(\WW)$
and the assumption $C_{\X}(\WW)<\infty$,
we have the fact that for any $\e>0$
there are a number $R$ and a pair consisting of an encoder $\vphi_n:\M_n\to\X^n$
and a decoder $\psi_n:\Y^n\to\M_n$ 
satisfying
\[
  C_{\X}(\WW)-\e\leq  \frac 1n\log|\M_n|\leq R<\infty
\]
and the conditions (\ref{eq:capacity-constraint}) and
(\ref{eq:capacity-error}).
Let $\cS$ be defined as
\[
  \cS_n\equiv\bigcup_{(x_1,\ldots,x_n)\in\vphi_n(\M_n)}\bigcup_{i=1}^n\{x_i\}.
\]
From the fact that $[1/n]\log|\M_n|\leq R<\infty$,
we have $|\cS_n|\leq n2^{nR}$, that is, $\cS_n$ is a finite set.
Furthermore, the image of $\vphi_n$ is a subset of $\cS_n^n$.
Let $q_n\equiv n2^{nR}$ and fix the set $\cS_n$ defined as above.
Then,
we have the fact that
\begin{align}
  C_{\X}(\WW)-\e
  &\leq
  C_{\X}^{q_n}(\WW)
  \notag
  \\
  &\leq
  \lim_{q\to\infty}C_{\X}^q(\WW)
\end{align}
for all $\e>0$ and all sufficiently large $n$,
where the first inequality
comes from the fact that $\cS_n\subset\X$
and $|\cS_n|\leq q_n$
and the last inequality comes from the fact that
$C_{\X}^q(\WW)$ is a non-decreasing function of $q$.
Hence we have
\[
  C_{\X}(\WW)\leq \lim_{q\to\infty}C_{\X}^q(\WW)+\e
\]
for all $\e>0$
and $C_{\X}(\WW)\leq \lim_{q\to\infty}C_{\X}^q(\WW)$ by letting
$n\to\infty$ and $\e\to0$.
\hfill\IEEEQED

\subsection{Proof of Theorem~\ref{thm:channel}}
\label{sec:proof-channel}

From (\ref{eq:channel-rR}),
we have the fact that there is $\e>0$ satisfying
\begin{align}
 r+R&<\uH(\XX)-\e.
 \label{eq:proof-channel-rR-e}
\end{align}

Let $\uT_X\subset\X^n$
be defined as
\[
  \uT_X
  \equiv
  \lrb{
    \xx:
    \frac 1n\log\frac 1{\mu_X(\xx)}
    \geq
    \uH(\XX)-\e
  }.
\]
For all $A$,
we have
\begin{align}
 &
 E_{\sfB\sfcc}\lrB{
  \sum_{\mm}
  \lrbar{
   \frac{\mu_X(\C_{A\sfB}(\sfcc,\mm))}
   {\mu_X(\C_A(\sfcc))}
   -
   \frac1
   {|\im\B|}
  }
 }
 \notag
 \\*
 &=
 E_{\sfB}\lrB{
  \sum_{\mm,\cc}
  \lrbar{
   \mu_X(\C_{A\sfB}(\cc,\mm))
   -
   \frac
   {\mu_X(\C_A(\cc))}
   {|\im\B|}
  }
 }
 \notag
 \\
 &\leq
 E_{\sfB}\lrB{
  \sum_{\mm,\cc}
  \lrbar{
   \mu_X(\C_{A\sfB}(\sfcc,\mm)\cap\uT_X)
   -
   \frac{\mu_X(\C_A(\cc)\cap\uT_X)}
   {|\im\B|}
  }
 }
 \notag
 \\*
 &\quad
 +
 E_{\sfB}\lrB{
  \sum_{\mm,\cc}\lrbar{
   \mu_X(\C_{A\sfB}(\sfcc,\mm)\cap[\uT_X]^c)
  }
 }
 +
 E_{\sfB}\lrB{
  \sum_{\mm,\cc}\lrbar{
   -\frac{\mu_X(\C_A(\cc)\cap[\uT_X]^c)}
   {|\im\B|}
  }
 }
 \notag
 \\
 \begin{split}
  &=
  E_{\sfB}\lrB{
   \sum_{\mm,\cc}
   \lrbar{
    \mu_X(\C_{A\sfB}(\cc,\mm)\cap\uT_X)
    -
    \frac{\mu_X(\C_A(\cc)\cap\uT_X)}
    {|\im\B|}
   }
  }
  \\*
  &\quad
  +
  E_{\sfB}\lrB{
   \sum_{\mm,\cc}
   \mu_X(\C_{A\sfB}(\cc,\mm)\cap[\uT_X]^c)
  }
  +
  E_{\sfB}\lrB{
   \sum_{\mm,\cc}
   \frac{\mu_X(\C_A(\cc)\cap[\uT_X]^c)}
   {|\im\B|}
  },
 \end{split}
 \label{eq:proof-channel-error2}
\end{align}
where the first equality comes from the fact
that
the distribution of the random variable $\sfcc$ is
$\{\mu_{X}(\C_A(\cc))\}_{\cc\in\im A}$,
the first inequality comes from
the triangular inequality
and 
the fact that
\begin{align*}
 \mu_{X}(\C_{AB}(\cc,\mm))
 &
 =\mu_{X}(\C_{AB}(\cc,\mm)\cap\uT_X)
 +\mu_{X}(\C_{AB}(\cc,\mm)\cap[\uT_X]^c)
 \\
 \mu_{X}(\C_A(\cc))
 &=
 \mu_{X}(\C_A(\cc)\cap\uT_X)
 +\mu_{X}(\C_A(\cc)\cap[\uT_X]^c),
\end{align*}
and the second equality comes from
the fact that
\begin{align*}
\mu_{X}(\C_{AB}(\cc,\mm)\cap[\uT_X]^c)&\geq0
\\
\mu_X(\C_A(\cc)\cap[\uT_X]^c)&\geq0.
\end{align*}
Since $\{C_{AB}(\cc,\mm)\}_{\cc,\mm}$ and
$\{C_A(\cc)\}_{\cc}$
form a partition,
the second and the third terms
on the right hand side of
(\ref{eq:proof-channel-error2})
are evaluated as
\begin{align}
  E_{\sfB}\lrB{
    \sum_{\mm,\cc}
    \mu_X(\C_{A\sfB}(\cc,\mm)\cap[\uT_X]^c)
  }
  &=
  \mu_{X}([\uT_X]^c)
  \\
  E_{\sfB}\lrB{
    \sum_{\mm,\cc}
    \frac{\mu_X(\C_A(\cc)\cap[\uT_X]^c)}
    {|\im\B|}
  }
  &=
  \mu_{X}([\uT_X]^c).
\end{align}
On the other hand, the first term on the right hand side of
(\ref{eq:proof-channel-error2})
is evaluated as
\begin{align}
 &
 E_{\sfB}\lrB{
  \sum_{\mm,\cc}
  \lrbar{
   \mu_X(\C_{A\sfB}(\cc,\mm)\cap\uT_X)
   -
   \frac{\mu_X(\C_A(\cc)\cap\uT_X)}
   {|\im\B|}
  }
 }
 \notag
 \\
 &=
 \sum_{\cc}
 \mu_X(\C_A(\cc)\cap\uT_X)
 E_{\sfB}\lrB{
  \sum_{\mm}
  \lrbar{
   \frac{\mu_X(\C_{\sfB}(\mm)\cap\C_A(\cc)\cap\uT_X)}
   {\mu_X(\C_A(\cc)\cap\uT_X)}
   -
   \frac1
   {|\im\B|}
  }
 }
 \notag
 \\
 &\leq
 \sum_{\cc}
 \mu_X(\C_A(\cc)\cap\uT_X)
 \sqrt{
  \alphaB-1
  +\frac{
   [\betaB+1]|\im\B|
   2^{-n[\uH(\XX)-\e]}}
  {\mu_X(\C_A(\cc)\cap\uT_X)}
 }
 \notag
 \\
 &\leq
 \mu_{X}(\uT_X)
 \sqrt{
  \sum_{\cc}
  \frac{\mu_X(\C_A(\cc)\cap\uT_X)}
  {\mu_{X}(\uT_X)}
  \lrB{
   \alphaB-1
   +
   \frac{
    [\betaB+1]|\im\B|
    2^{-n[\uH(\XX)-\e]}}
   {\mu_X(\C_A(\cc)\cap\uT_X)}
  }
 }
 \notag
 \\
 &=
 \mu_{X}(\uT_X)
 \sqrt{
  \alphaB-1
  +
  \frac{[\betaB+1]|\im A||\im\B|2^{-n[\uH(\XX)-\e]}}
  {\mu_{X}(\uT_X)}
 }
 \notag
 \\*
 &\leq
 \sqrt{\alphaB-1+[\betaB+1]2^{-n[\uH(\XX)-r-R-\e]}}
 \label{eq:proof-channel-error3}
\end{align}
where the fist inequality comes from Lemma~\ref{lem:BCP}
and the fact that
$\mu_X(\xx)\leq 2^{-n[\uH(\XX)-\e]}$
for all $\xx\in\uT_X$,
the second inequality comes from the Jensen inequality,
and the last inequality comes from (\ref{eq:r}),
(\ref{eq:R}), and the fact that $\mu_X(\uT_X)\leq 1$.
Then we have
\begin{align}
  &
  E_{\sfB\sfcc}\lrB{\Error(A,\sfB,\sfcc)}
  \notag
  \\*
  &=
  E_{\sfB\sfcc}\lrB{
   \sum_{\substack{
     \mm:
     \\
     \mu_{X^n}(\C_{A\sfB}(\sfcc,\mm))=0
   }}
   \frac1
   {|\im\B|}
   +
   \sum_{\substack{
     \mm,\xx,\yy:\\
     \mu_{X^n}(\C_{A\sfB}(\sfcc,\mm))>0\\
     \xx\in\C_{A\sfB}(\sfcc,\mm)\\
     \xx_{A}(\sfcc|\yy)\neq\xx
   }}
   \frac{\mu_{XY}(\xx,\yy)}
   {|\im\B|\mu_X(\C_{A\sfB}(\sfcc,\mm))}
  }
  \notag
  \\
  &=
  E_{\sfB\sfcc}\left[
   \sum_{\substack{
     \mm:
     \\
     \mu_{X^n}(\C_{A\sfB}(\sfcc,\mm))=0
   }}
   \!\!\!\!\!\!\!\!\!
   \frac1
   {|\im\B|}
   +
   \sum_{\substack{
     \mm,\xx,\yy:\\
     \mu_{X^n}(\C_{A\sfB}(\sfcc,\mm))>0\\
     \xx\in\C_{A\sfB}(\sfcc,\mm)\\
     \xx_A(\sfcc|\yy)\neq\xx
   }}
   \!\!\!\!\!\!\!\!\!
   \mu_{XY}(\xx,\yy)
   \lrB{
    \frac1{\mu_X(\C_A(\cc))}
    +
    \frac1
    {|\im\B|\mu_X(\C_{A\sfB}(\sfcc,\mm))}
    -
    \frac1{\mu_X(\C_A(\cc))}
   }
  \right]
  \notag
  \\
  &\leq
  \sum_{\substack{
    \xx,\yy:\\
    \xx_A(A\xx|\yy)\neq\xx
  }}
  \mu_{XY}(\xx,\yy)
  +
  E_{\sfB\sfcc}\lrB{
   \sum_{\mm}
   \lrbar{
    \frac{\mu_X(\C_{A\sfB}(\sfcc,\mm))}
    {\mu_X(\C_A(\cc))}
    -
    \frac1{|\im\B|}
   }
  }
  \notag
  \\
  &\leq
  \Error(A)
  +
  \sqrt{
   \alphaB-1
   +
   [\betaB+1]2^{-n[\uH(\XX)-r-R-\e]}
  }
  +2\mu_X([\uT_X]^c)
  \label{eq:proof-channel-error}
\end{align}
where
the first inequality comes from the fact that
\begin{align}
 E_{\sfB\sfcc}\lrB{
  \sum_{\substack{
    \mm,\xx,\yy:\\
    \xx\in\C_{A\sfB}(\sfcc,\mm)\\
    \xx_A(\sfcc|\yy)\neq\xx
  }}
  \frac{\mu_{XY}(\xx,\yy)}
  {\mu_X(\C_A(\cc))}
 }
 &=
 E_{\sfB}\lrB{
  \sum_{\substack{
    \cc,\mm,\xx,\yy:\\
    \xx\in\C_{A\sfB}(\sfcc,\mm)\\
    \xx_A(A\xx|\yy)\neq\xx
  }}
  \mu_{XY}(\xx,\yy)
 }
 \notag
 \\
 &=
 \sum_{\xx,\yy}
 \mu_{XY}(\xx,\yy)\chi(\xx_A(A\xx|\yy)\neq\xx)
\end{align}
and
\begin{align}
  &
  \sum_{\substack{
      \mm,\xx,\yy:\\
      \mu_X(\C_{AB}(\cc,\mm))>0
      \\
      \xx\in\C_{AB}(\cc,\mm)
  }}
  \mu_{XY}(\xx,\yy)
  \lrB{
    \frac1
    {|\im\B|\mu_X(\C_{AB}(\cc,\mm))}
    -
    \frac1{\mu_X(\C_A(\cc))}
  }
  \notag
  \\
  &\leq
  \sum_{\substack{
      \mm:\\
      \mu_X(\C_{AB}(\cc,\mm))>0
  }}
  \mu_X(\C_{AB}(\cc,\mm))
  \lrbar{
    \frac1
    {|\im\B|\mu_X(\C_{AB}(\cc,\mm))}
    -
    \frac1{\mu_X(\C_A(\cc))}
  }
  \notag
  \\
  &=
  \sum_{\substack{
      \mm:\\
      \mu_X(\C_{AB}(\cc,\mm))>0
  }}
  \lrbar{
    \frac{\mu_X(\C_{AB}(\cc,\mm))}
    {\mu_X(\C_A(\cc))}
    -
    \frac1
    {|\im\B|}
  }
  \notag
  \\
  &=
  \sum_{\mm}
  \lrbar{
    \frac{\mu_X(\C_{AB}(\cc,\mm))}
    {\mu_X(\C_A(\cc))}
    -
    \frac1
    {|\im\B|}
  }
  -
  \sum_{\substack{
      \mm:\\
      \mu_X(\C_{AB}(\cc,\mm))=0
  }}
  \frac1
  {|\im\B|},
\end{align}
and the second inequality comes from
(\ref{eq:sw}) and
(\ref{eq:proof-channel-error2})--(\ref{eq:proof-channel-error3}).
From 
(\ref{eq:alpha}),
(\ref{eq:beta}),
(\ref{eq:proof-channel-rR-e}),
(\ref{eq:proof-channel-error}),
and the fact that
$\mu_X([\uT_X]^c)\to0$
as $n\to\infty$,
we have the fact that there is a pair consisting of a function $B\in\B$
and a vector $\cc\in\im A$ that satisfy (\ref{eq:channel-error}).
\hfill\IEEEQED

\subsection{Proof of Lemma~\ref{lem:ex}}
\label{sec:proof-ex}
Since 
$\pA\lrsb{\lrb{
      A: 
      \bt(\xx)\in\hcH\cup\{\bt(\zero)\}
      \ \text{for all}\ \xx\in\C_A(\zero)
}}$
depends only on $\hcH$
and $\pA\lrsb{\lrb{A: A\xx=\zero}}$ depends on $\xx$
only through the type $\bt(\xx)$,
we have the fact that
$\tpA\lrsb{\lrb{A: A\xx=\zero}}$
also depends on $\xx$ only through the type $\bt(\xx)$.
We use the following lemma.
\begin{lem}[{\cite[Lemma~6]{CRNG},\cite[Theorem~1]{HASH-BC}}]
\label{lem:hash-linear}
Let $(\A,\pA)$
be an ensemble of matrices.
We assume that \\
$\pA\lrsb{\lrb{A: A\xx=\zero}}$
depends on $\xx$ only through the type $\bt(\xx)$.
Let $(\alphaA,\betaA)$ be defined by
(\ref{eq:alpha-linear}) and (\ref{eq:beta-linear}).
Then $(\A,\pA)$ satisfies (\ref{eq:hash-cr}).
\end{lem}

Now, we prove the lemma.
Let $\chi(\cdot)$ be defined by (\ref{eq:chi}).
From the fact that
$\cH\setminus\hcH$ and $\hcH\cup\{\bt(\zero)\}$ are
the disjoint union of the set of all types of length $n$,
we have
\begin{align}
 \pA\lrsb{\lrb{
   A: 
   \bt(\xx)\in\hcH\cup\{\bt(\zero)\}
   \ \text{for all}\ \xx\in\C_A(\zero)
 }}
 &=
 1-
 \pA\lrsb{\lrb{
   A: 
   \exists\xx\in\C_A(\zero)
   \ \text{s.t.}\ \bt(\xx)\in\cH\setminus\hcH
 }}
 \notag
 \\
 &\geq
 1
 -\sum_{\bt\in\cH\setminus\hcH}
 \sum_{\xx}\pA\lrsb{\lrb{A: A\xx=\zero,\bt(\xx)=\bt}}
 \notag
 \\
 &=
 1
 -\sum_{\bt\in\cH\setminus\hcH}
 \sum_{\xx}
 \sum_A\pA(A)
 \chi(A\xx=\zero,\bt(\xx)=\bt)
 \notag
 \\
 &=
 1
 -\sum_{\bt\in\cH\setminus\hcH}
 \sum_A\pA(A)
 \sum_{\xx}
 \chi(A\xx=\zero,\bt(\xx)=\bt)
 \notag
 \\
 &=
 1
 -\sum_{\bt\in\cH\setminus\hcH}
 \sum_A\pA(A)|\{\xx\in\X^n: A\xx=\zero, \bt(\xx)=\bt\}|
 \notag
 \\
 &=
 1
 -\sum_{\bt\in\cH\setminus\hcH}
 S(\pA,\bt)
 \notag
 \\
 &=
 1-\betaA
\end{align}
and
\[
  \frac1
  {\pA\lrsb{\lrb{
      A: 
      \bt(\xx)\in\hcH\cup\{\bt(\zero)\}
      \ \text{for all}\ \xx\in\C_A(\zero)
  }}}
  \leq
  \frac1{1-\betaA}
\]
for all sufficiently large $n$ satisfying $\betaA<1$.
We have
\begin{align}
  \tpA(A)
  &\leq
  \frac{\pA(A)}
  {\pA\lrsb{\lrb{
	A: 
	\bt(\xx)\in\hcH\cup\{\bt(\zero)\}
	\ \text{for all}\ \xx\in\C_A(\zero)
    }}
  }
  \notag
  \\
  &\leq
  \frac{\pA(A)}
  {1-\betaA}
  \label{eq:tpA}
\end{align}
for all $A$ and all sufficiently large $n$.
Let $\talphaA^*$ be defined as
\[
  \talphaA^*\equiv
  \frac{|\im\A|}{|\X|^l}\cdot\max_{\bt\in \hcH}
  \frac{S(\tpA,\bt)}{S(p_{\osfA},\bt)}.
\]
Then we have the relation
\begin{align}
  \talphaA^*
  &=
  \frac{|\im\A|}{|\X|^l}
  \max_{\bt\in \hcH}\sum_A
  \frac{\tpA(A)|\{\xx: A\xx=\zero,\bt(\xx)=\bt\}|}
  {S(p_{\osfA},\bt)}
  \notag
  \\
  &\leq
  \frac{|\im\A|}{|\X|^l}
  \max_{\bt\in \hcH}\sum_A
  \frac{\pA(A)|\{\xx: A\xx=\zero,\bt(\xx)=\bt\}|}
  {\lrB{1-\betaA}S(p_{\osfA},\bt)}
  \notag
  \\
  &=
  \frac{|\im\A|}{\lrB{1-\betaA}|\X|^l}
  \cdot\max_{\bt\in \hcH}
  \frac{S(\pA,\bt)}{S(p_{\osfA},\bt)}
  \notag
  \\
  &=
  \frac{\alphaA}{1-\betaA}
  \notag
  \\
  &=
  \talphaA
  \label{eq:proof-ex-alpha}
\end{align}
for all sufficiently large $n$,
where the inequality comes from (\ref{eq:tpA}).

When $\bt\in\cH\setminus\hcH$ and
$|\{\xx: A\xx=\zero, \bt(\xx)=\bt\}|>0$,
there is $\xx\in\C_A(\zero)$ such that
$\bt(\xx)\in\cH\setminus\hcH$.
Then, from the definition of $\tpA$, 
we have $\tpA(A)=0$
when $\bt\in\cH\setminus\hcH$ and
$|\{\xx: A\xx=\zero, \bt(\xx)=\bt\}|>0$.
This implies that
\begin{align}
 \sum_{\bt\in\cH\setminus\hcH}S(\tpA,\bt)
 &=
 \sum_{\bt\in\cH\setminus\hcH}
 \sum_A\tpA(A)|\{\xx: A\xx=\zero, \bt(\xx)=\bt\}|
 \notag
 \\
 &=0.
 \label{eq:proof-ex-beta}
\end{align}
Then we have
\begin{align}
 \sum_{\substack{
   \xx'\in\X^n\setminus\{\xx\}:
   \\
   \tpA(\{A: A\xx = A\xx'\})>\frac{\talphaA}{|\im\A|}
 }}
 \tpA\lrsb{\lrb{A: A\xx = A\xx'}}
 &\leq
 \sum_{\substack{
   \xx'\in\X^n\setminus\{\xx\}:
   \\
   \tpA(\{A: A\xx = A\xx'\})>\frac{\talphaA^*}{|\im\A|}
 }}
 \tpA\lrsb{\lrb{A: A\xx = A\xx'}}
 \notag
 \\
 &\leq
 \sum_{\bt\in\cH\setminus\hcH}S(\tpA,\bt)
 \notag
 \\
 &=0
 \label{eq:proof-ex-hash}
\end{align}
for all sufficiently large $n$ satisfying $\betaA<1$,
where
the first inequality comes from (\ref{eq:proof-ex-alpha}),
the second inequality is obtained by applying
Lemma~\ref{lem:hash-linear} to the ensemble $(\A,\tpA)$
and $\talphaA^*$,
and the equality comes from (\ref{eq:proof-ex-beta}).

We have
\begin{align}
 \limsupn\frac 1n\log\talphaA(n)
 &=
 \limsupn\frac 1n\log\frac{\alphaA(n)}{1-\betaA(n)}
 \notag
 \\
 &\leq
 \limsupn\frac 1n\log\alphaA(n)
 +
 \limsupn\frac 1n\log\frac 1{1-\betaA(n)}
 \notag
 \\
 &=0,
 \label{eq:proof-ex-CR1}
\end{align}
where the last equality comes from 
(\ref{eq:alpha-ex}) and (\ref{eq:beta-ex}).
Then, from (\ref{eq:proof-ex-hash}) and (\ref{eq:proof-ex-CR1}),
we have the fact that $(\bcA,\tbpA)$ satisfies
the $(\taalphaA,\zero)$-collision-resistance property.
Inequality (\ref{eq:CRPex}) is shown directly by
applying Lemma~\ref{lem:CRP} to $(\bcA,\tbpA)$.
\hfill\IEEEQED

\subsection{Proof of Theorem~\ref{thm:si}}
\label{sec:proof-si}

From (\ref{eq:sir}), we have the fact that
there is $\e>0$ satisfying
\begin{align}
  r
  &>\oH(\XX|\YY)+\e.
  \label{eq:proof-si-re}
\end{align}
Let $\oT_{X|Y}\subset\X^n\times\Y^n$
be defined as
\begin{align*}
  \oT_{X|Y}
  &\equiv
  \lrb{
    (\xx,\yy):
    \frac 1n\log\frac 1{\mu_{X|Y}(\xx|\yy)} \leq \oH(\XX|\YY)+\e
  }.
  \label{eq:channel-TXY}
\end{align*}

Assume that $(\xx,\yy)\in\oT_{X|Y}$ and $\xx_A(A\xx|\yy)\neq \xx$.
Then we have the fact that there is
$\xx'\in\C_A(A\xx)$ such that $\xx'\neq\xx$ and
\[
  \mu_{X|Y}(\xx'|\yy)\geq
  \mu_{X|Y}(\xx|\yy)
  \geq 2^{-n[\oH(\XX|\YY)+\e]}.
\]
This implies that
$\lrB{\oT_{X|Y}(\yy)\setminus\{\xx\}}\cap\C_A(A\xx)\neq\emptyset$,
where 
$\oT_{X|Y}(\yy)\equiv\{\xx: (\xx,\yy)\in\oT_{X|Y}\}$.
We have
\begin{align}
 E_{\sfA}\lrB{
  \chi(\xx_{\sfA}(\sfA\xx|\yy)\neq\xx)
 }
 &\leq
 p_{\sfA}\lrsb{\lrb{A:
   [\oT_{X|Y}(\yy)\setminus\{\xx\}]\cap\C_A(A\xx)\neq\emptyset
 }}
 \notag
 \\*
 &\leq
 \frac{|\oT_{X|Y}(\yy)|\alphaA{}}
 {|\im\A|}
 +\betaA{}
 \notag
 \\
 &\leq
 2^{-n[r-\oH(\XX|\YY)-\e]}\alphaA{}
 +\betaA{}
 \label{eq:proof-si-error-sub}
\end{align}
for all $(\xx,\yy)\in\oT_{X|Y}$,
where $\chi(\cdot)$ is defined by (\ref{eq:chi}),
the second inequality comes from Lemma~\ref{lem:CRP},
and the third inequality comes from
(\ref{eq:sir}) and
the fact that
$|\oT_{X|Y}(\yy)|\leq 2^{n[\oH(\XX|\YY)+\e]}$.

We have the fact that
\begin{align}
 &
 E_{\sfA}\lrB{\Error(\sfA)}
 \notag
 \\*
 &=
 E_{\sfA}\lrB{
  \sum_{\xx,\yy}
  \mu_{XY}(\xx,\yy)
  \chi(\xx_{\sfA}(\sfA\xx|\yy)\neq\xx)
 }
 \notag
 \\
 &=
 \sum_{(\xx,\yy)\in\oT_{X|Y}}
 \mu_{XY}(\xx,\yy)
 E_{\sfA}\lrB{
  \chi(\xx_{\sfA}(\sfA\xx|\yy)\neq\xx)
 }
 +
 \sum_{(\xx,\yy)\notin\oT_{X|Y}}
 \mu_{XY}(\xx,\yy)
 E_{\sfA}\lrB{
  \chi(\xx_{\sfA}(\sfA\xx|\yy)\neq\xx)
 }
 \notag
 \\
 &\leq
 2^{-n[r-\oH(\XX|\YY)-\e]}\alphaA{}
 +\betaA{}
 +\mu_{XY}([\oT_{X|Y}]^c),
 \label{eq:proof-channel-error1}
\end{align}
where the last inequality comes from (\ref{eq:proof-si-error-sub}).
From (\ref{eq:alpha-cr}), (\ref{eq:beta-cr}), (\ref{eq:proof-si-re})
and the fact that $\mu_{XY}([\uT_{XY}]^c)\to0$ as $n\to\infty$,
we have the fact that there is a function $A\in\A$
satisfying $\Error(A)\leq\delta$
for all $\delta>0$ and all sufficiently large $n$.
\hfill\IEEEQED

\appendix

\subsection*{Error Probability of Stochastic Decision}

This appendix reviews the result of \cite{SDECODING},
which investigates the error probability
of the stochastic decision.
It should be noted that stochastic decoding
is a stochastic decision in the context of a coding scheme.

Let $\U$ and $\V$ be the alphabets of random variable $U$ and $V$, respectively.
We assume that the joint distribution $p_{UV}$ of $(U,V)$ is known.

Let us assume the situation where a decoder make a
stochastic decision
of the invisible state $U$ after the observation $V$.
We use a random number generator $\hU\in\U$
after observing $V$
and let $\hU$ be a decision (guess) about the state $U$.
Formally, we generate $\hU$ subject to the conditional distribution
$q_{\hU|V}(\cdot|V)$ on $\U$ depending on an observation $V$
and let an output be a decision of
$U$, where $U$ and $\hU$ are conditionally independent
for a given $V$, that is, $U\markov V\markov\hU$ forms a Markov chain.
The joint distribution $p_{UV\hU}$ of $(U,V,\hU)$
is given as
\[
 p_{UV\hU}(u,v,\hu)=q_{\hU|V}(\hu|v)p_{U|V}(u|v)p_V(v).
\]
Let us call $q_{\hU|V}$ a {\em stochastic decision rule}.
As a special case, when $q_{\hU|V}$
is given by using a function $f:\U\to\V$ and is defined as
\begin{equation}
 q_{\hU|V}(\hu|v)
 =
 \begin{cases}
  1 &\text{if}\ \hu=f(v)
  \\
  0 &\text{if}\ \hu\neq f(v),
 \end{cases}
 \label{eq:deterministic}
\end{equation}
we call $q_{\hU|V}$ or $f$ a {\em deterministic decision rule}.
It should be noted that the maximum a posteriori decision rule is
deterministic.

Let $\chi$ be a support function defined by (\ref{eq:chi}).
Then the error probability $\Error(q_{\hU|V})$ of
a (stochastic) decision rule $q_{\hU|V}$ is given as
\begin{align}
 \Error(q_{\hU|V})
 &=
 \sum_{v}p_V(v)
 \sum_{u}
 p_{U|V}(u|v)
 \sum_{\hu}
 q_{\hU|V}(\hu|v)
 \chi(\hu\neq u)
 \notag
 \\
 &=
 \sum_{v}p_V(v)
 \sum_{u}
 p_{U|V}(u|v)[1-q_{\hU|V}(u|v)].
 \label{eq:error-random}
\end{align}
In the last equality,
$1-q_{\hU|V}(u|v)$ corresponds to the error probability
of the decision rule $q_{\hU|V}$
after the observation $v\in\V$,
and $\Error(q_{\hU|V})$ corresponds to the average of this error probability.
When $q_{\hU|V}$ is defined by using $f:\V\to\U$ and (\ref{eq:deterministic}),
the decision error probability $\Error(f)$ of a deterministic decision
rule $f$ is given as
\begin{align}
 \Error(f)
 &\equiv
 \sum_{v}p_V(v)\sum_{u}p_{U|V}(u|v)\chi(f(v)\neq u)
 \notag
 \\
 &=
 \sum_{v}p_V(v)[1-p_{U|V}(f(v)|v)].
 \label{eq:error-f}
\end{align}
It should be noted that
the right hand side of the first equality
can be derived directly from
(\ref{eq:error-random}) and the fact that
\begin{align}
 q_{\hU|V}(u|v)
 &=\chi(f(v)=u)
 \notag
 \\
 &=1-\chi(f(v)\neq u).
\end{align}
That is,
we have $\Error(f)=\Error(q_{\hU|V})$
when $f$ and $q_{\hU|V}$ satisfy (\ref{eq:deterministic}).

It is well-known fact (see \cite[Lemma 2]{SDECODING}) that
an optimal strategy for guessing the state $U$
is finding $\hu$ which maximize
the conditional probability $p_{U|V}(\hu|v)$
depending on a given observation $v$.
Formally, by taking $\hu$ that maximizes
$p_{U|V}(\hu|v)$ for each $v\in\V$,
we can define the function $\fMAP:\V\to\U$ as
\begin{align}
 \fMAP(v)&
 \equiv\arg\max_{\hu} p_{U|V}(\hu|v)
 \label{eq:map}
 \\
 &=\arg\max_{\hu} p_{UV}(\hu,v).
 \label{eq:ml}
\end{align}
We call (\ref{eq:map}) and (\ref{eq:ml})
a {\em maximum a posteriori decoder}
and a {\em maximum likelihoood decoder}, respectively.
It should be noted that the discussion
does not depend on the choice of states with the same maximum
probability.

Here, let us consider the case $q_{\hU|V}(\hu|v)=p_{U|V}(\hu|v)$
for all $(\hu,v)$, that is,
we make a stochastic decision with the conditional distribution
$p_{U|V}$ of a state $U$ for a given observation $V$.
It should be noted that the joint distribution $p_{UV\hU}$ of
$(U,V,\hU)$ is given as
\[
 p_{UV\hU}(u,v,\hu)
 =
 p_{U|V}(\hu|v)p_{U|V}(u|v)p_V(v).
\]
We call this type of decision rule a
{\em stochastic decision with the a posteriori distribution}.

We have the following lemma.
\begin{lem}[{\cite[Eq.~(29)]{CH67}\cite[Lemma 3]{SDECODING}}]
\label{lem:sdecoding}
Let $(U,V)$ be a pair consisting of a state $U$ and an observation $V$
and $p_{UV}$ be the joint distribution of $(U,V)$.
When we make a stochastic decision with $p_{U|V}$,
the decision error probability of this rule
is at most twice the decision error probability
of the maximum a posteriori decision rule $\fMAP$.
That is, we have
\begin{equation*}
 \Error(p_{U|V})\leq 2\Error(\fMAP).
 \label{eq:random-fmap}
\end{equation*}
\end{lem}
\begin{IEEEproof}
In this proof, we assume that $\U$ and $\V$ are finite sets.
It should be noted that the result does not change when
$\V$ is an infinite/continuous set, where 
the summation should be replaced with the integral.
We have
\begin{align}
 \Error(p_{U|V})
 &=
 \sum_{v}p_V(v)
 \sum_{u}
 p_{U|V}(u|v)
 [1-p_{U|V}(u|v)]
 \notag
 \\
 &=
 \sum_{v}p_V(v)
 \lrB{1-\sum_{u}p_{U|V}(u|v)^2}
 \notag
 \\
 &\leq
 \sum_{v}p_V(v)
 \lrB{1-p_{U|V}(\fMAP(v)|v)^2}
 \notag
 \\
 &=
 \sum_{v}p_V(v)
 [1-p_{U|V}(\fMAP(v)|v)]
 [1+p_{U|V}(\fMAP(v)|v)]
 \notag
 \\
 &\leq
 2\sum_{v}p_V(v)
 [1-p_{U|V}(\fMAP(v)|v)]
 \notag
 \\
 &=
 2\Error(\fMAP),
 \label{eq:proof-random-deterministic}
\end{align}
where the second inequality comes from the fact that
$p_{U|V}(f(v)|v)\leq 1$ and the fourth equality comes from (\ref{eq:error-f}).
\end{IEEEproof}


\begin{thebibliography}{99}
\addcontentsline{toc}{chapter}{Bibliography}
\bibitem{AC98}
R.\ Ahlswede and I.\ Csisz\'{a}r,
``Common randomness in information theory and cryptography
--- Part II: CR capacity,''
{\it IEEE Trans.\ Inform.\ Theory},
vol.~IT-44, no.~1,
pp.~225--240, Jan.\ 1998.
\bibitem{A09}
E.\ Ar\i{}kan,
``Channel polarization: a method for constructing capacity-achieving
codes for symmetric binary-input memoryless channels,''
{\it IEEE Trans.\ Inform.\ Theory},
vol.~IT-55, no.~7, July 2009.
\bibitem{BM01}
J. Bajcsy and P. Mitran,
``Coding for the Slepian-Wolf problem with turbo codes,''
{\it Proc.\ 2001 IEEE Globecom}, pp.~1400--1404, 2001.
\bibitem{BB04}
A.\ Bennatan and D.\ Burshtein,
``On the application of LDPC codes to arbitrary discrete-memoryless
channels,''
{\it IEEE Trans.\ Inform.\ Theory},
vol.~IT-50, no.~3, pp.~417--438, Mar.~2004.
\bibitem{BGT93}
C.\ Berrou, A.\ Glavieux, and P.\ Thitimajshima,
``Near Shannon limit error correcting codes and decoding: turbo codes,''
{\it Proc.\ Int.\ Conf.\ Commun.},
Geneva, Switzerland, May 23--26, 1993, pp.~1064--1070.
\bibitem{B87}
R.\ E.\ Blahut,
{\em Principles and Practice of Information Thoery},
Addison-Welsey Publishing. Co., 1987.
\bibitem{CW}
J.\ L.\ Carter and M.\ N.\ Wegman,
``Universal classes of hash functions,''
{\it J.\ Comput.\ Syst.\ Sci.}, vol.\ 18, pp.\ 143--154, 1979.
\bibitem{CHJLY09}
J.\ Chen, D.K.\ He, A.\ Jagmohan, L.\ A.\ Lastras-Monta\~{n}o,
and E.H.\ Yang,
``On the linear codebook-level duality between Slepian-Wolf coding and
channel coding,''
{\it IEEE Trans. Inform Theory},
vol.~IT-55, no.~12, pp.~5575--5590, Dec.~2009.
\bibitem{C75}
T.\ M.\ Cover,
``A proof of the data compression theorem of Slepian and Wolf for ergodic sources,''
{\it IEEE Trans. Inform Theory},
vol.~IT-21, no.~2, pp.~226--228, Mar.\ 1975.
\bibitem{CH67}
T.\ M.\ Cover and P.\ E.\ Hart,
``Nearest neighbor pattern classification,''
{\it IEEE Trans.\ Inform.\ Theory},
vol.\ IT-13, no.\ 1, pp.\ 21--27, Jan.\ 1967.
\bibitem{CSI82}
I.\ Csisz\'{a}r,
``Linear codes for sources and source networks:
				Error exponents, universal coding,''
{\it IEEE Trans.\ Inform.\ Theory},
vol.\ IT-28, no.\ 4, pp.\ 585--592, Jul.\ 1982.
\bibitem{CK11}
I.\ Csisz\'{a}r and J.\ K\"{o}rner,
{\it Information Theory: Coding Theorems for Discrete Memoryless Systems
	2nd Ed.},
Cambridge University Press, 2011.
\bibitem{E55}
P.\ Elias,
``Coding for noisy channels,''
{\it IRE Convention Record}, Part 4, pp.~37--46, 1955.
\bibitem{EM05}
U.\ Erez and G.\ Miller,
``The ML decoding performance of LDPC ensembles over
				$\boldsymbol{Z}_q$,''
{\it IEEE Trans.\ Inform.\ Theory},
vol.~IT-51, no.~5, pp.~1871--1879, May 2005.
\bibitem{GA68}
R.\ G.\ Gallager,
{\it Information Theory and Reliable Communication},
John Wiley \& Sons, Inc., 1968.
\bibitem{GA62}
R.\ G.\ Gallager,
{\it Low Density Parity Check Codes},
Cambridge, MA:M.I.T Press, 1963.
\bibitem{GZ01}
J.\ Garcia-Frias and Y.\ Zhao,
``Compression of correlated binary memoryless sources using turbo codes,''
{\it IEEE Comm.\ Letters}, vol.~5, no.\ 10, pp.\ 417--419, Oct.\ 2001.
\bibitem{HV93}
T.S.\ Han and S.\ Verd\'u,
``Approximation theory of output statistics,''
{\it IEEE Trans.\ Inform.\ Theory},
vol.~IT-39, no.~3, pp.~752--772, May 1993.
\bibitem{HAN}
T.S.\ Han,
{\it Information-Spectrum Methods in Information Theory},
Springer, 2003.
\bibitem{HY13}
J.\ Honda and H.\ Yamamoto,
``Polar coding without alphabet extension for asymmetric models,''
{\it IEEE Trans.\ Inform.\ Theory},
vol.~IT-59, no.~12, pp.~7829--7838, Dec.\ 2013.
\bibitem{HKU09}
N.\ Hussami, S.\ B.\ Korada, and R.\ Urbanke,
``Performance of polar codes for channel and source coding,''
{\it Proc.\ 2009 IEEE Int.\ Symp.\ Inform.\ Theory},
Seoul, Korea, Jun.~28--Jul.~3, 2009, pp.~1488--1492.
\bibitem{IZ89}
R.\ Impagliazzo and D.\ Zuckerman,
``How to recycle random bits,''
{\it 30th IEEE Symp. Fund. Computer Sci.},
Oct.~30--Nov.~1, 1989, pp.~248--253.
\bibitem{MB01}
G.\ Miller and D.\ Burshtein,
``Bounds on the maximum-likelihood decoding error probability of
low-density parity-check codes,''
{\it IEEE Trans.\ Inform.\ Theory},
vol.\ IT-47, no.\ 7, pp.\ 2696--2710, Nov.\ 2001.
\bibitem{MM08}
S.\ Miyake and J.\ Muramatsu,
``A construction of channel code, JSCC and universal code
				for discrete memoryless channels using sparse matrices,''
{\it IEICE Trans.\ Fundamentals}, vol.~E92-A, no.~9,
        pp.~2333--2344, Sep.\ 2009.
\bibitem{MK95}
S.\ Miyake and F.\ Kanaya,
``Coding theorems on correlated general sources,''
{\it IEICE Trans.\ Fundamentals}, vol.\ E78-A, no.\ 9,
pp.\ 1063--1070, Sept.\ 1995.
\bibitem{CRNG}
J.\ Muramatsu,
``Channel coding and lossy source coding using a generator of
constrained random numbers,''
{\it IEEE Trans.\ Inform.\ Theory},
vol.\ IT-60, no.\ 5, pp.\ 2667--2686, May 2014.
\bibitem{CRNG-VLOSSY}
J.\ Muramatsu,
``Variable-length lossy source code using a constrained-random-number generator,''
{\it IEEE Trans.\ Inform.\ Theory},
	vol.~IT-61, no.~6, Jun.~2015.
\bibitem{HASH}
J.\ Muramatsu and S.\ Miyake,
``Hash property and coding theorems for sparse matrices and
                                maximal-likelihood coding,''
{\it IEEE Trans.\ Inform.\ Theory}, vol.\ IT-56, no. 5, pp.\ 2143--2167,
        May 2010.
 Corrections: vol.\ IT-56, no.\ 9, p.\ 4762, Sep.\ 2010,
vol.\ IT-59, no.\ 10, pp.\ 6952--6953, Oct.\ 2013.
\bibitem{HASH-BC}
J.\ Muramatsu and S.\ Miyake,
``Construction of
Slepian-Wolf source code and broadcast channel code
based on hash property,''
available at {\tt arXiv:1006.5271[cs.IT]}, 2010.
\bibitem{ISIT2010}
J.\ Muramatsu and S.\ Miyake,
``Construction of broadcast channel code
based on hash property,''
{\it Proc.\ 2010 IEEE Int.\ Symp.\ Inform.\ Theory},
Austin, U.S.A., June 13--18, 2010, pp.\ 575--579.
\bibitem{ISIT2011a}
J.\ Muramatsu and S.\ Miyake,
``Construction of strongly secure wiretap channel code
based on hash property,''
{\it Proc.\ 2011 IEEE Int.\ Symp.\ Inform.\ Theory},
St.\ Petersburg, Russia, Jul.\ 31--Aug.\ 5, 2011, pp.\ 612--616.
\bibitem{COUNTABLELIMIT}
J.\ Muramatsu and S.\ Miyake,
``Fundamental limits are achievable with  countable alphabet,''
{\it Proc.\ 2016 Int.\ Symp.\ Inform.\ Theory and its
 Applicat.},
Monterey, California, USA, Oct.~30--Nov.~2, 2016, pp.~573--577.
\bibitem{SDECODING}
J.\ Muramatsu and S.\ Miyake,
``On the error probability of stochastic decision and stochastic
decoding,''
to appear in {\it Proc.\ 2017 IEEE Int.\ Symp.\ Inform. Theory},
Aachen, Germany, June 25--30, 2017.
Extended version is available at
{\tt arXiv:1701.04950[cs.IT]}, 2017.
\bibitem{SWLDPC}
J.\ Muramatsu, T.\ Uyematsu, and T.\ Wadayama,
``Low density parity check matrices for coding of correlated sources,''
{\it IEEE Trans.\ Inform.\ Theory}, vol.\ IT-51, no.\ 10,
pp.\ 3645--3653, Oct.\ 2005.
\bibitem{M02}
T.\ Murayama,
``Statistical mechanics of data compression theorem,''
{\it J.\ Phys.\ A: Math.\ Gen.},
vol.\ 35: L95L100, 2002.
\bibitem{RR11}
J.\ M.\ Renes and R.\ Renner,
``Noisy channel coding via privacy amplification
and information reconciliation,''
{\it IEEE Trans.\ Inform.\ Theory},
vol.~IT-57, no.~11, pp.~7377--7385, Nov.\ 2011.
\bibitem{SPR02}
D.\ Schonberg, S.\ S.\ Pradhan, and K.\ Ramchandran,
``LDPC codes can approach the Slepian-Wolf bound for general binary
        sources,''
{\it 40th Annual Allerton Conference on Communication, Control, and
        Computing,}
Allerton House, Monticello, Illinois, Oct.\ 2002.
\bibitem{SW73}
D.\ Slepian and J.\ K.\ Wolf,
``Noiseless coding of correlated information sources,''
{\it IEEE Trans.\ Inform.\ Theory}, vol.~IT-19, no.~4,
pp.~471--480, Jul.~1973.
\bibitem{S71}
J.\ G.\ Smith,
``The information capacity of amplitude- and variance-constrained
scalar Gaussian Channels,''
{\it Inf.\ Control}, vol.\ 18, pp.\ 203--219, 1971.
\bibitem{SV93}
Y.\ Steinberg and S.\ Verd\'u,
``Channel simulation and conding with side information,''
{\it IEEE Trans.\ Inform.\ Theory}, vol.~IT-40, no.~3, pp.~634--646, May 1994.
\bibitem{VH94}
S. Verd\'u and T.S.\ Han,
``A general formula for channel capacity,''
{\it IEEE Trans. Inform. Theory}, vol.~IT-40, no.~4, pp.~1147--1157, Jul.~1994.
\bibitem{W74}
A.\ D.\ Wyner,
``Recent results in the Shannon theory,''
{\it IEEE Trans.\ Inform.\ Theory}, vol.~IT-20, no.~1, pp.~2--10, Jan.~1974.
\bibitem{YAG12}
M.\ H.\ Yassaee, M.\ R.\ Aref, and A.\ Gohari,
``Achievability proof via output statistics of random binning,''
{\it IEEE Trans.\ Inform.\ Theory},
vol.~IT-60, no.~11, pp.~6760-6786, Nov.~2014.
\end{thebibliography}
\end{document}